\documentclass[12pt]{article}
\usepackage{amsmath,amssymb,graphicx,physics,url}
\usepackage{setspace}
\usepackage{geometry}
\title{\textbf{Physical Analysis of a Reported Missile--``Orb'' Interaction in 2024:\\
Momentum Constraints, Atmospheric Drag, Sensor Artifacts, and Theoretical Caution}}

\author{
Mauricio Barbi \\
University of Regina, Department of Physics \\
3737 Wascana Parkway, Regina, SK, S4S 0A2, Canada \\
\texttt{barbi@uregina.ca}
}

\date{}

\begin{document}
\maketitle
\onehalfspacing

\begin{abstract}
A widely circulated infrared video from 2024 appears to show an air-to-surface missile engaging a small luminous ``orb''-like object, producing debris and an apparent deflection of the missile's trajectory. This paper presents a comprehensive physics-based analysis of the event. Using classical mechanics, fluid dynamics, and imaging-system geometry, we examine momentum transfer, debris behavior, gimbal-induced optical distortions, and the physical plausibility of various interpretations. A balloon-like object remains the most conservative explanation compatible with known physics. However, modern physics is incomplete: the extreme weakness of gravity relative to the gauge forces, the existence of dark matter, dark energy, neutrino masses, and the possibility of extra spatial dimensions indicate that the Standard Model is not the final description of nature. For this reason, anomalous observations should not be dismissed outright as ``impossible.'' Nevertheless, extraordinary claims require extraordinary evidence, and the present video lacks the multi-sensor, verified, and calibrated data required to support exotic interpretations. The aim is a balanced scientific evaluation based on current knowledge while maintaining appropriate theoretical humility.
\end{abstract}

\section{Introduction}

A thermal infrared video released publicly in 2024 \cite{DoD-UAP2024} and later shown during a U.S. Congressional hearing in 2025 \cite{HouseUAP2025} appears to depict an AGM-114-class missile approaching, striking, and possibly interacting with a small luminous object. Interpretations range from a balloon impact to exotic technological hypotheses. This paper evaluates the event using established physics while acknowledging the incompleteness of fundamental theory. Known physics provides strong constraints, but modern physics contains open questions—such as the hierarchy problem, dark sector phenomena, and potential extra dimensions—that caution against absolute claims of impossibility.

Our goal is to identify what the video \emph{requires}, \emph{allows}, and does \emph{not allow} under known physics, and what additional data would be needed to evaluate alternative interpretations.

\section{Limits of Known Physics and the Need for Theoretical Caution}

Classical mechanics, hydrodynamics, and optics are remarkably predictive across wide regimes. Yet major gaps remain in modern physics:

\begin{itemize}
    \item Gravity is $\sim 10^{40}$ times weaker than the strong interaction.
    \item Gauge couplings unify at high energies, whereas gravity does not.
    \item Neutrino masses require new physics beyond the Standard Model.
    \item Dark matter and dark energy dominate the energy density of the universe.
    \item Baryon asymmetry requires additional CP-violating processes.
\end{itemize}

These facts strongly suggest that known physics is incomplete. However, in regimes where classical hydrodynamics, Newtonian mechanics, and optical physics have been extensively validated—such as missile flight through atmosphere—departures from known behavior would require extraordinary evidence. We therefore use established physics as the baseline for interpretation, while avoiding categorical dismissal of alternatives.

\section{Kinematic Considerations and Momentum Transfer}

Assuming a missile mass $m_m \approx 45\,\mathrm{kg}$ and velocity $v_m \approx 400\,\mathrm{m/s}$ \cite{FAS-Hellfire}, its momentum is:

\begin{equation}
p_m = m_m v_m \approx 1.8\times 10^4~\mathrm{kg\,m/s}.
\label{eq:missile_momentum}
\end{equation}

A balloon-like object with mass $m_b \sim 0.2$–$1~\mathrm{kg}$ \cite{NOAA-Balloons} has momentum at least two orders of magnitude smaller. Meteorological balloons drift with the surrounding air mass, and typical atmospheric wind speeds at operational altitudes range from approximately $5~\mathrm{m/s}$ to $20~\mathrm{m/s}$; here, to remain conservative, we also consider an extreme upper-bound case in which the balloon moves at $50~\mathrm{m/s}$.

Even in an inelastic collision, the maximum lateral deflection angle is:

\begin{equation}
\theta_{\max} \approx \tan^{-1}\left(\frac{p_b}{p_m}\right) \approx 0.32^\circ,
\label{eq:deflection_angle}
\end{equation}

consistent with classical collision theory \cite{Goldstein,Thornton}. Larger apparent deflections must therefore originate from sensor geometry rather than true missile dynamics.

\section{Stress, Fragmentation, and Debris Dynamics}

The apparent debris produced at the impact point may arise from:

\begin{itemize}
    \item tearing of thin balloon envelopes,
    \item reflective panels or lightweight internal structures,
    \item localized stress concentrations during rupture.
\end{itemize}

These fragments have negligible ballistic inertia and quickly enter a drag-dominated regime.

\subsection{Drag-Dominated Behavior of Light Fragments}

For debris mass $m_f \sim 10^{-3}~\mathrm{kg}$ and area $A \sim 10^{-2}~\mathrm{m^2}$ \cite{Yoo}, the drag force is:

\begin{equation}
F_D = \frac{1}{2}\rho C_d A v_{\mathrm{rel}}^2,
\label{eq:drag}
\end{equation}

with $\rho \approx 1.2~\mathrm{kg/m^3}$ and $C_d \sim 1.3$ \cite{Anderson,Hoerner}. Resulting decelerations exceed $10^4$–$10^5~\mathrm{m/s^2}$, giving characteristic damping times:

\begin{equation}
\tau \approx \frac{v}{a} \sim 10^{-3}~\mathrm{s}.
\label{eq:damping_time}
\end{equation}

Thus, debris rapidly loses forward momentum and is carried by ambient airflow, consistent with the observed behavior. In addition to the small fragments, the remaining bulk of the balloon envelope is also strongly influenced by aerodynamic drag. Once ruptured, the balloon structure has a large effective area-to-mass ratio and therefore rapidly tends to align its motion with the surrounding air stream. Any transient forward momentum acquired in the missile’s direction is quickly overcome by drag forces and by advection in the local flow. This behaviour is consistent with the visual impression that the main body of the object and its debris are carried along by the air rather than continuing with a significant component of motion along the missile’s original trajectory.

\section{Energy Considerations}

The missile’s kinetic energy is:

\begin{equation}
K_m = \frac{1}{2} m_m v_m^2 \approx 3.6\times 10^{6}~\mathrm{J}.
\label{eq:missile_ke_again}
\end{equation}

\subsection*{Elastic and Total Mechanical Energy of the Balloon}

The elastic strain energy stored in a stretched thin film is given by the standard expression from linear elasticity \cite{Timoshenko1961}:

\begin{equation}
U = \frac{1}{2} E \,\epsilon^{2} V ,
\label{eq:elastic_energy}
\end{equation}

where $E$ is Young’s modulus, $\epsilon$ the strain, and $V = A t$ the material volume.

Measured balloon-film parameters \cite{Ray1971} give:

\[
E \sim 1\text{--}10~\mathrm{MPa}, \qquad
\epsilon \sim 0.3\text{--}1.0, \qquad
A \sim 3~\mathrm{m^2}, \qquad
t \sim 50~\mu\mathrm{m}.
\]

Substituting these values into Eq.~\eqref{eq:elastic_energy} yields:

\begin{equation}
U \sim 10\text{--}100~\mathrm{J}.
\label{eq:balloon_energy_range}
\end{equation}

Even in an extreme upper-bound case with $m_b = 1~\mathrm{kg}$ and $v_b = 50~\mathrm{m/s}$, the balloon's kinetic energy,

\begin{equation}
K_b = \frac{1}{2} m_b v_b^2 \approx 1.3\times 10^{3}~\mathrm{J},
\label{eq:balloon_ke}
\end{equation}

remains roughly three orders of magnitude smaller than $K_m$. The total mechanical energy available in the balloon, $K_b + U$, is therefore negligible compared to the missile's energy and cannot supply any dynamically significant impulse to alter the missile's trajectory.

\section{Imaging Geometry and Gimballed Sensor Artifacts}

The infrared footage was acquired by a gimballed auto-tracking sensor \cite{Holst,Hassan}. Such systems introduce well-known artifacts \cite{ONeil} such as:

\begin{itemize}
    \item apparent curvature of straight trajectories,
    \item false accelerations from non-inertial frame rotation,
    \item sudden ``direction changes'' from gimbal recentering,
    \item horizon-loss contrast drop,
    \item rotation of IR glare interpreted as object rotation.
\end{itemize}

A number of mundane, everyday situations illustrate how rotation of the imaging frame can create apparent changes in an object's trajectory. For example, when a person films a moving car while panning a handheld camera, the car often appears to follow a curved or abruptly changing path whenever the camera adjusts its rotation speed or recenters the subject in the frame. A similar effect is familiar to passengers filming the ground from a commercial airliner: when the aircraft banks or changes pitch, the landscape appears to “slide” or rotate even though the terrain is fixed, demonstrating how line-of-sight rotation produces apparent motion. An even closer analogue occurs when tracking a moving object using a fluid-head tripod, where small adjustments in pan--tilt motion introduce apparent accelerations or bends in the subject’s trajectory despite the target moving linearly. These everyday examples illustrate that apparent curvature, acceleration, or directional changes in infrared footage can arise naturally from rotation of the sensor’s line of sight rather than from any anomalous behavior of the tracked object.

\section{Alternative Interpretations and Theoretical Neutrality}

The event could in principle be interpreted as:

\begin{itemize}
    \item a balloon or lightweight decoy,
    \item a reconnaissance or spoofing device,
    \item sensor or gimbal artifact,
    \item undisclosed advanced technology.
\end{itemize}

Only the first three are required by the available evidence.  
The latter requires data not present.

\section{Discussion}

The video lacks:

\begin{itemize}
    \item sensor metadata,
    \item platform telemetry,
    \item range and scale calibration,
    \item multi-sensor correlation,
    \item radar or environmental data,
    \item chain-of-custody verification.
\end{itemize}

Thus, the IR footage alone is insufficient to support extraordinary claims. All observed features remain fully compatible with known physics.

\section{Context from U.S. UAP Hearings and Evidence Limitations}

U.S. Congressional UAP hearings (2023--2025) presented several ambiguous videos, including this missile--orb clip. All shared the same limitations: no sensor metadata, no radar corroboration, no multi-angle imagery, susceptibility to optical artifacts, and absence of verified provenance. These hearings underscore the necessity of multi-sensor, calibrated, chain-of-custody--verified data before drawing extraordinary conclusions.

\section{Conclusions}

Using classical mechanics, drag physics, and imaging geometry, we find:

\begin{itemize}
    \item a balloon-like target is the most conservative explanation,
    \item no element of the video requires exotic physics,
    \item apparent missile deflection is consistent with gimbal artifacts,
    \item debris behavior matches drag-dominated dynamics.
\end{itemize}

The UAP-hearing context further reinforces that the evidence is insufficient for extraordinary interpretations. While modern physics is incomplete and alternative possibilities cannot be categorically ruled out, the available data strongly favor conventional explanations. A definitive evaluation requires multi-sensor, high-fidelity, chain-of-custody--verified data.

\section*{Disclosure}

Artificial Intelligence (ChatGPT) was used as a research-assistance tool to help organize the document, explore publicly available literature, clarify wording, and polish the manuscript. Mundane examples of possible gimballed sensor artifacts were elaborated with the help of ChatGPT. All scientific developments, interpretations, and conclusions were introduced and fully validated by the human author.


\end{document}